\documentclass[aps,prl,floatfix,reprint]{revtex4-1}
\usepackage{epsfig}
\usepackage{amsmath}
\usepackage{amsfonts}
\usepackage{amssymb}
\usepackage{mathptmx}
\usepackage{eucal}
\usepackage{bm}

\begin{document}

\title{Effect of rf electromagnetic irradiation on current-voltage characteristics of wide superconducting films}

\author{I.~V.~Zolochevskii}
\affiliation{B. Verkin Institute for Low Temperature Physics and Engineering, 61103 Kharkov, Ukraine}
\author{A.~V.~Terekhov}
\affiliation{B. Verkin Institute for Low Temperature Physics and Engineering, 61103 Kharkov, Ukraine}
\author{E.~V.~Bezuglyi}
\affiliation{B. Verkin Institute for Low Temperature Physics and Engineering, 61103 Kharkov, Ukraine}
\author{L.~A.~Ishchenko}
\affiliation{B. Verkin Institute for Low Temperature Physics and Engineering, 61103 Kharkov, Ukraine}
\author{E.~V.~Khristenko}
\affiliation{B. Verkin Institute for Low Temperature Physics and Engineering, 61103 Kharkov, Ukraine}

\begin{abstract}
The effect of the electromagnetic irradiation on the current-voltage characteristics (IVCs) of a wide superconducting film is experimentally investigated. In contrast to the microwave field (several GHz) which suppresses vortex resistivity, the action of the rf field (tens of MHz) leads to significant expansion of the linear part of the IVC due to rapid suppression of the critical current with a slower change in the upper stability limit of the vortex state where the phase-slip lines (PSLs) occur. With an increase in the rf power, the stepwise structure of the IVC associated with the PSLs becomes smoothed and eventually disappears. A model of the IVC in the adiabatic regime is proposed, which explains the effects of smoothing of the voltage steps and the suppression of the critical current.
\end{abstract}

\maketitle

\section{1. Introduction} \vspace{-1mm}
The mechanisms of the influence of electromagnetic radiation on a superconducting film can be divided into two groups: bolometric and non-bolometric. Bolometric mechanisms caused by overheating of the electronic system are well-studied \cite{Bol1,Bol2}. Investigations of non-bolometric (non-equilibrium) effects of electromagnetic irradiation of superconducting films are still relevant today, since they reveal microscopic mechanisms of the electromagnetic response and relaxation of the order parameter and quasiparticles which are important for efficient operation of superconducting radiation sensors. Historically, traditional objects for studying these effects were narrow films (superconducting channels) in the frequency range of several GHz, mainly due to the interest in the phenomena of superconductivity stimulation by the microwave field. In this paper, we study the effect of a high-frequency field in the relatively little explored rf range (tens of MHz) on the resistive state of wide superconducting tin films.

The films were fabricated according to the original technology \cite{Zol2005}, which ensures high quality of the film side edges and a uniform thickness. This has been confirmed by good agreement of the experimentally measured values and temperature dependences of their critical currents \cite{Zol2005,Zol2010} with the results of the Aslamazov-Lempitzky theory \cite{AL,Blok} for the defect-free films. The films were deposited on the substrates of optically polished monocrystalline quartz, which ensures an efficient heat removal \cite{Kaplan} and the absence of overheating.

The resistive current state of such films is realized via two mechanisms:  viscous motion of the vortex lattice and formation of the phase-slip lines (PSLs), which are activated sequentially as the transport current increases. First, when the current reaches a certain critical value $I_c$ at which the current density at the edges of the film approaches the depairing value $j_c^{GL}$ of the Ginzburg-Landau theory, then the edge energy barrier for the vortex penetration disappears, and vortices with opposite signs start to penetrate into the film from different edges \cite{AL,LO}. This leads to the emergence of the vortex resistive part of the current-voltage characteristic (IVC). Annihilation of the vortices at the center of the film generates a peak in the current density \cite{AL} which has been confirmed by the numerical simulation \cite{Bez2015} and visualized in the experiment \cite{peak1,peak2}. With a sufficiently large value $I_m$ of the transport current, the height of this peak reaches $j_c^{GL}$, and the vortex lattice becomes unstable \cite{AL}, which leads to the formation of PSLs and the voltage steps on the IVC.

Fabrication of the homogeneous wide films with smooth edges is quite a challenge; historically, films with a large number of defects were first investigated. For instance, when studying the rf field absorption by the vortices in Pb$_{0.83}$In$_{0.17}$ and Nb$_{0.95}$Ta$_{0.05}$ films with a large number of pinning centers \cite{Gitt}, it was found that the alternating field of certain frequencies above $3.9-15$ MHz causes the appearance of vortex resistivity and the corresponding increase in the absorbed power due to depinning of vortices. However, in our early studies of the effect of the microwave field ($1$--$15$ GHz) on the vortex resistive state in the films with a small number of defects \cite{Zol2009}, the opposite effect compared to \cite{Gitt} was found: suppression of the vortex resistivity by the electromagnetic field. It was shown that with increasing microwave power $P$,
the resistive vortex part of the IVC decreases and completely disappears at $P \geq  0.4P_c$ ($P_c$ is the power at which the critical current vanishes). The question of the influence of a lower frequency (similar to that used in \cite{Gitt}) on the vortex resistivity of high-quality films remained open, which caused our interest in the study of the resistive state of the films irradiated with the rf electromagnetic wave.
\vspace{-1mm}

\section{2. Experiment}

In our experiments, we used several Sn films 18--42 $\mu$m wide, about $90$ $\mu$m long, and 120--330 nm thick; the results were similar for all samples. When measuring the IVC by the four-probe method, the samples were placed in a double screen made from the annealed permalloy which reduces the magnetic field in the region of the sample to the values $H_\bot = 7\times 10^{-4}$ Oe, $H_\| = 6.5\times 10^{-3}$ Oe. The electric component of the electromagnetic field is directed parallel to the transport current. Since the film length was less than 1\% of the minimum wavelength, the high-frequency current $I_f \propto \sqrt P$ induced by the electromagnetic field is almost constant along the film length.

\begin{figure}[ht]
\centerline{\epsfxsize=8.5cm\epsffile{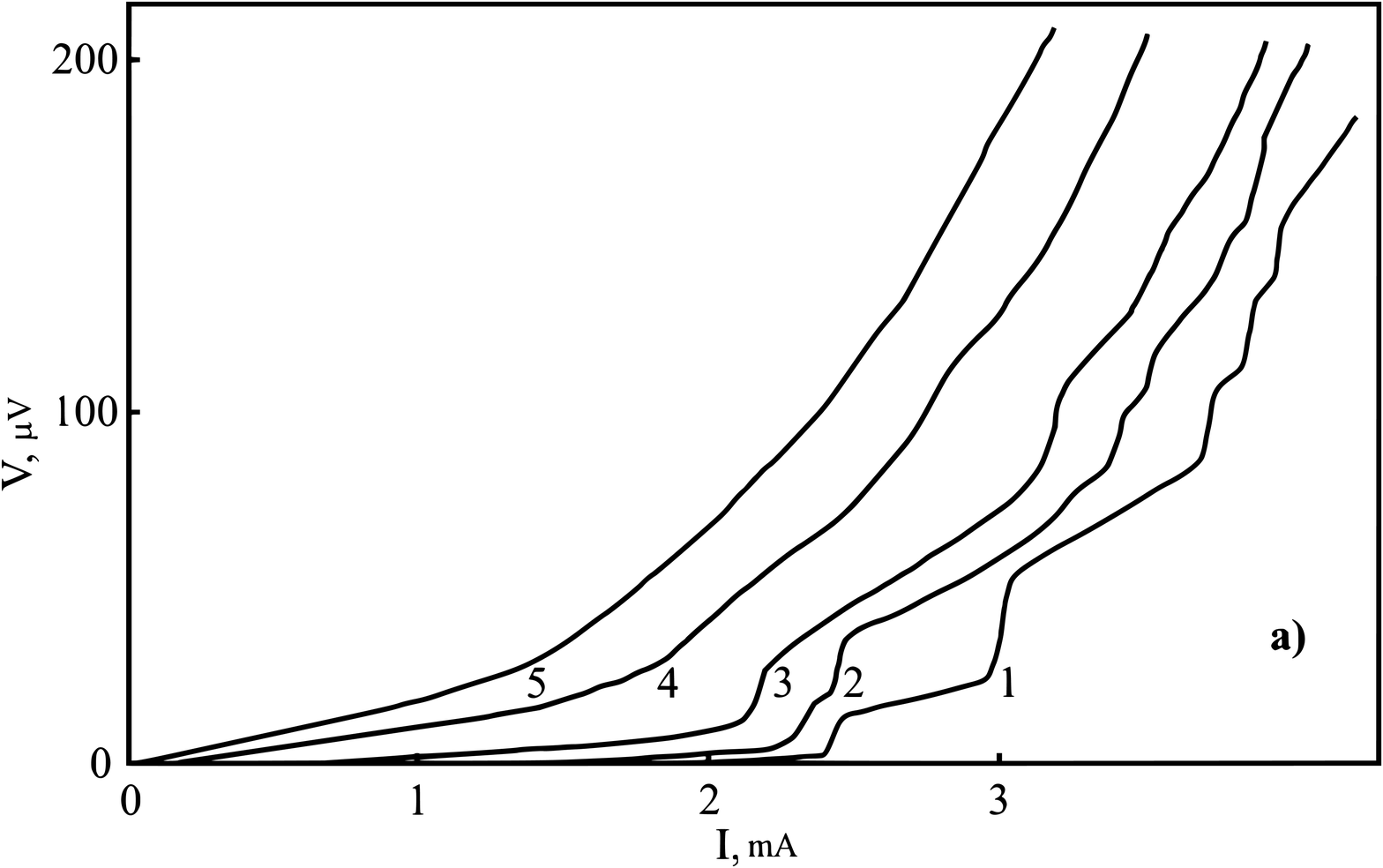}}
\vspace{5mm}
\centerline{\epsfxsize=8.5cm\epsffile{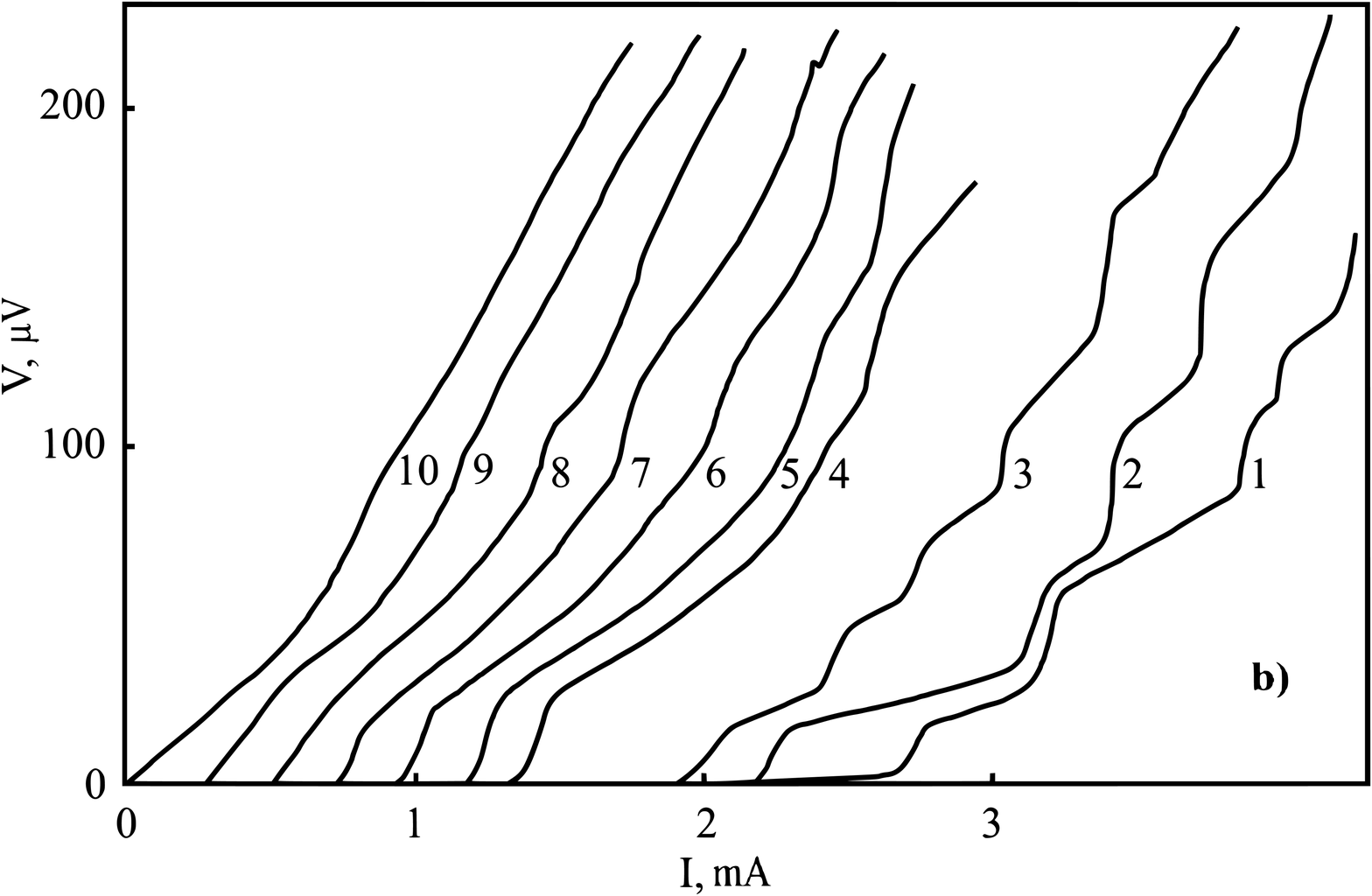}}
\caption{Experimental IVCs of a wide (42 $\mu$m) Sn film at various electromagnetic power levels (a) in the rf band, $f = 35.5$ MHz, $T = 3.754$ K  and (b) in the microwave band, $f = 5.56$ GHz, $T = 3.744$ K. For the IVC No.1, the irradiation power is $P = 0$, and for others it grows with their numbers.} \vspace{-5mm}
\label{exper}
\end{figure}

Fig. 1 shows the families of the IVCs of the sample Sn4w (width 42 $\mu$m, length 92 $\mu$m, thickness 120 nm, the temperature of the superconducting transition $T_c = 3.795$ K, and the normal resistance $0.14$ $\Omega$ at $T = 4.2$ K), measured at different power levels at frequencies of $35.5$ MHz (a) and $5.56$ GHz (b). For the IVCs marked by $1$, the irradiation power is zero, and for the others it grows with their sequence numbers. As seen in Fig. 1b, under the action of microwave irradiation ($f=5.56$ GHz), the film resistivity, caused by the motion of the vortices (a small initial linear part of the IVC No. 1), decreases and then disappears at a certain power level, in accordance with earlier results \cite{Zol2009}; simulaneously, the current $I_m$ of occurrence of the first PSL decreases and finally turns to zero.

The effect of an alternating field of the rf band ($f=35.5$ MHz, Fig. 1a) is radically different: the critical current $I_c$ decreases significantly faster with increasing power than $I_m$. As a result, the initial part of the IVC considerably extends, its slope increases, and most of this part remains linear. In addition, with an increase in the rf power, the voltage jumps caused by the emergence of PSLs, becomes smoothed and eventually disappear, while under microwave irradiation this effect is markedly weaker. A similar smoothing effect in the rf region ($75$ MHz) has been previously detected in narrow channels, where the IVC was formed by phase-slip centers in the absence of the vortex resistivity \cite{Zol1986}.

\section{3. Model and Discussion}

A fundamental role in the formation of the response of a superconductor to an alternating field is played by the relaxation time $\tau_\Delta$ of the order parameter $\Delta(t)$ \cite{Pals}, which is related to the time $\tau_\epsilon$ of the energy relaxation of electrons as $\tau_\Delta \approx 1.2 \tau_\epsilon (1-T/T_c)^{-1/2}$ near the critical temperature. For our samples and temperatures, $\tau_\epsilon \approx 4.3\times 10^{-10}$ s \cite{Zol2005a}, and the order parameter relaxation time is $\tau_\Delta \approx 4.2\times 10^{-9}$ s. Thus, in the microwave range, the relaxation time $\tau_\Delta$ turns out to be large compared with the period of the electromagnetic wave ($\approx 2\times 10^{-10}$ s), thus, the order parameter $\Delta(t)$ can not significantly change during the wave period and experiences only weak oscillations around its average value $\bar\Delta$ which is determined by the average intensity of the alternating field \cite{Bez1987}. For this reason, the effect of the alternating field on the stepwise IVC structure in the high-frequency limit $\omega\tau_\Delta \gg 1$ is mainly reduced to a decrease in the onset current $I_m$ due to suppression of $\bar\Delta$. A possible cause of the disappearance of the vortex resistivity may be rapid oscillatory motion of the vortices caused by the microwave current with a small period. This prevents formation of new vortices at the edges of the film, which, according to \cite{AL}, requires much longer time of the order of $\tau_\Delta$ (see a detailed discussion in \cite{Zol2009}).

In the rf range, when the relaxation times of the order parameter and the formation of vortices are small compared with the period of an electromagnetic wave ($\approx 3\times 10^{-8}$ s), the order parameter and the vortex structure follow the changes in the field almost adiabatically. In this case, one can assume that at each moment of the time $t$, the film is in a locally quasi-static steady state with the total current $I_{tot}(t) = I+I_f \sin\omega t$. Thus, as soon as the sum of the transport current $I$ and the amplitude $I_f$ of the alternating current exceeds the critical current $I_c(0)$ of formation of the vortices in zero electromagnetic field, the vortex resistivity occurs in the film. The corresponding critical current
\begin{equation}
I_c(P) = I_c(0)-I_f
\end{equation}
decreases rather quickly with increasing power, in accordance with the experiment. We note that the simple relation (1) makes it possible to estimate the alternating current excited in the film for each IVC in Fig. 1a, that is inaccessible for a direct measurement. In particular, the critical current of the film should vanish at the value of $I_f$ equal to the critical current $I_c(0)$ in the absence of pumping, which in our experiment is approximately $2.1$ mA.

Similar considerations can be applied to describe the shape of the IVC, using the fact that the instantaneous voltage value $V_t(I,P)$ in the adiabatic limit $\omega\tau_\Delta \ll 1$ is close to the value $V(I_{tot}(t),0)$ at the IVC in the absence of irradiation, which corresponds to the instantaneous total current $I_{tot}(t)$. Since the constant voltage $V(I,P)$ measured in the experiment is the average of $V_t(I,P)$ over the period $T_f$ of the field oscillations,
\begin{equation} \label{IVCmodel}
V(I,P) = \int_0^{T_f} \frac{dt}{T_f} V(I+I_f \sin\omega t,0).
\end{equation}
then each point on the IVC of the film under irradiation (2) actually represents the result of a certain averaging of the IVC in the absence of pumping, $V(I,0)$, over the vicinity of the transport current $I$ of width $2I_f$. In particular, this explains the experimentally observed smoothing of the voltage jumps, which enhances with increasing power.

\begin{figure}[!h]
\centerline{\epsfxsize=8.5cm\epsffile{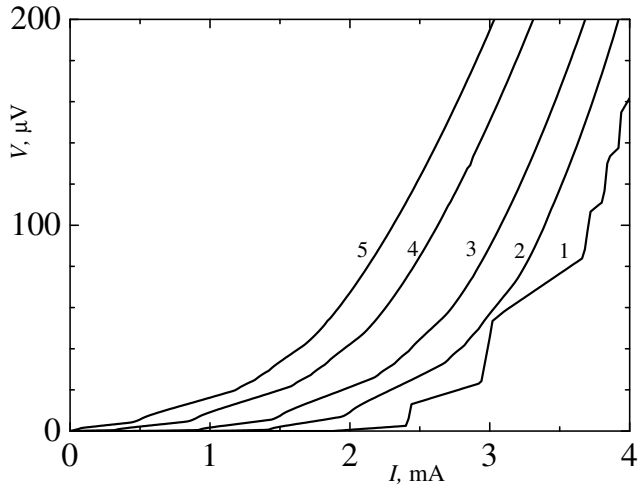}}
\caption{Calculated IVCs (curves 2--5) obtained from the formula (2) using the experimental IVC measured at $P = 0$ (curve 1). The numbering of the calculated IVCs corresponds to different values of alternating current: 2 -- $1$ mA, 3 -- $1.5$ mA, 4 -- $2$ mA, 5 -- $2.5$ mA.} 
\label{model}
\end{figure}

The results of numerical calculations of the IVCs, obtained by substitution of the digitized experimental IVC at $P=0$ to the formula (2) at various values of $I_f$, are presented in Fig.~2. They reveal a good qualitative similarity with the results of the experiment, except for the fact that disappearance of the voltage jumps with increasing power occurs much faster than in the experiment. Also, the initial sections of the IVC in this model turn out to be noticeably nonlinear, in contrast to the linear experimental ones. Among the possible reasons for these differences, we first of all mention the insufficiently low frequency of irradiation (in the experiment, the relaxation parameter value $\omega\tau_\Delta \approx 0.9$ most likely corresponds to the transition regime to the adiabatic limit $\omega\tau_\Delta \ll 1$), as well as the existence of other characteristic times of the system. For example, the vortex flight time $\tau$ across the film, according to estimates in \cite{Zol2009}, may exceed $\tau_\Delta$ by an order of magnitude and even more. Since the adiabaticity condition for this time, $\omega\tau_\Delta \ll 1$, is not satisfied in our experiment, the assumption of a locally steady state may be violated, therefore the quantitative applicability of the model requires usage of a lower-frequency radiation.

\section{4. Conclusion}

In conclusion, we have performed a comparative study of the influence of the microwave and the rf electromagnetic fields on the resistive state of a wide high-quality Sn film. In contrast to the microwave field which leads to suppression of the vortex resistivity on the IVC, the rf field irradiation results in a significant expansion of the linear part of the IVC. This can be interpreted as a more rapid suppression of the critical current of vortex penetration into the film compared to the upper limit of the instability of the vortex state, above which phase-slip lines occur. With an increase in the power of rf irradiation, the stepwise structure of the IVC, caused by the onsets of the PSLs, is smoothed and eventually disappears. A model of the IVC of the film in the low-frequency (adiabatic) regime, as a result of averaging the static IVC over the oscillations of the alternating current component, is proposed. This model qualitatively explains the effects of smoothing of the voltage steps and the rapid decrease of the critical current, and allows one to estimate the amplitude of the alternating current excited in the film .

\end{document}